\documentclass[conference]{IEEEtran}

\AtBeginDocument{%
  }

\usepackage{nameref}
\usepackage{caption}
\usepackage{subcaption}
\usepackage{multirow}
\usepackage{listings}
\usepackage{hyperref}
\usepackage{graphicx}
\usepackage{comment}
\usepackage{booktabs}
\usepackage{balance}

\newcommand{\code}[1]{\texttt{#1}}%

\begin{document}

\title{Measuring Weather Effects and Link Quality Dynamics in LEO Satellite Networks}

\author{Clemens Lottermoser$^1$ \quad Simon Damm$^1$ \quad Stefan Schmid$^{1,3,4}$\\
$^1$TU Berlin \quad $^2$TU Munich \quad  $^3$Fraunhofer SIT  \quad  $^4$Weizenbaum Institute}

\newcommand{\foothref}[1]{\footnote{\href{#1}{#1}}}
\newcommand{\Autoref}[1]{%
  \begingroup%
  \def\chapterautorefname{Chapter}%
  \def\sectionautorefname{Section}%
  \def\subsectionautorefname{Section}%
  \def\subsubsectionautorefname{Section}%
  \autoref{#1}%
  \endgroup%
}

\maketitle

\begin{abstract}
This paper presents an empirical study of dynamic factors affecting link quality in Low Earth Orbit (LEO) satellite communications, using Starlink as a case study. Over 56 days, 112 high-quality meteorological measurements in mostly 1-min intervals, co-located with a user terminal, were collected, alongside frequent network performance data. Cloud characteristics were estimated using professional weather instruments such as a ceilometer, microwave radiometer, and vision-language model on sky images. Our results show that general cloud presence does not significantly impact throughput or latency. The impact of cloud coverage rather depends on the presence of liquid water in the atmosphere, quantified by liquid water path (LWP), which correlates with notable download throughput reductions (up to 60 MBit/s), especially during rain. Upload and latency were largely unaffected. Analysis of the evolving satellite network revealed that newer satellite hardware and infrastructural upgrades also contributed to performance increases during the experiment period. These findings highlight atmospheric liquid water as the key weather-related factor affecting link quality and underscore the influence of network changes over time.
\end{abstract}

\section{Introduction}\label{sec:introduction-leo-satellite-networks}

In the last 5 years, Low Earth Orbit (LEO) satellites have become a new transformative telecommunications component. Unlike traditional geostationary (GEO) satellites located at 35,786km, LEO satellites have a much closer orbit of 300km to 2000km. This proximity significantly reduces latency and offers new opportunities for global connectivity, especially in previously unserved areas. 

Among current LEO Satellite providers such as \textit{Eutelsat OneWeb}, \textit{Telesat}, \textit{Amazon Project Kuiper}, \textit{SpaceX's Starlink} is the largest and the only one with commercial offerings to private users. As of April 19, 2025, Starlink has 7,207 active satellites in orbit out of a total of 10,513 active satellites operating there.
These LEO Satellite Networks (LSNs) aim to provide direct high-speed, low-latency internet access to end users worldwide. In our UDP-based performance tests with the current Starlink Standard Kit Gen. 3 (Rev. 4) in central Europe, we achieve maximum download speeds of 490 MBit/s and 150 MBit/s upload speeds, with a minimum latency of 21ms (cf. \Autoref{sec:network_performance}). 

LSNs offer several improvements over traditional broadband connections, such as global coverage and resilience to ground infrastructure failure. 
However, performance is not immune to external influences. One of the most critical factors, especially for real-time applications, is link quality, which can be influenced by various dynamic factors. This includes atmospheric influences (e.g., temperature, rain, cloud cover, ionospheric activity), satellite motion, and local topography and obstructions \cite{maNetworkCharacteristicsLEO2023}. LEO satellites move at a speed of over 27,000 km/h \cite{bhattacherjeeNetworkTopologyDesign2019} and a handover to a new satellite is performed every 27 seconds on average in our findings (cf. \Autoref{sec:starlink_routing_schedule}). 
Understanding the root causes of these influencing factors is essential to optimizing the performance and reliability of satellites, especially as they are increasingly incorporated into critical infrastructure and emergency-response systems. 

While previous research has explored weather impacts on LEO satellite performance, existing studies have only partially addressed these influences. In particular, prior work has mostly relied on a smaller set of weather variables and commodity sensors, which makes it harder to disentangle the effects of rain, cloud cover, and other forms of atmospheric liquid water.

This paper is motivated by the opportunity to fill these gaps with a novel approach: by leveraging new satellite detection techniques presented by \cite{ahangarpourTrajectorybasedServingSatellite2024} and integrating a richer set of professional meteorological instruments and higher-resolution network measurements, we can systematically analyze how satellite infrastructure, e.g., different satellite versions, responds to fine-grained weather phenomena, including both cloud cover and liquid water path. Because of the ever-changing network characteristics of the Starlink system, challenges of this work include the compensation of varying measurements due to noticeable hourly drifts (e.g., day and nighttime) and global fluctuations in network performance. This enables a more precise, real-world understanding of the dynamic factors shaping LEO satellite performance and supports both the further development of these networks and the optimization of user experience.

Specifically, this paper makes three \textbf{contributions}. 
\emph{First}, it is demonstrated that despite the absence of direct data from Starlink, it is viable to estimate connected satellites and evaluate their connection influences. This is achieved by integrating Two-Line Element (TLE) data, representing satellite locations, and Starlink diagnostic data, which enables the comparison of different versions of Starlink satellites and the impact on performance from handovers and distance to the dish. Furthermore, implementing a ping analysis (every 10ms) to the Starlink gateway provides insights into the internal mechanisms of the Starlink infrastructure.

\emph{Second}, this work proposes alternative methodologies for investigating cloud cover, employing a sky camera and a robust large-language model (LLM) pipeline with vision capabilities (VLM) to formulate statements about cloud cover, thereby markedly reducing the challenges associated with investigating these factors. This facilitates the formulation of precise statements regarding cloud coverage and its influence on LEO satellite communication, even in the absence of precipitation.

\emph{Third}, we conduct a statistical analysis of over 115 weather and network parameters and their influences on the current Starlink Gen. 3 Dish. This enables the differentiation of various influences such as rain, cloud cover, other atmospheric conditions, and internal network data. The results show that these factors have measurable effects on network throughput, with variations in download speeds reaching up to 60 MBit/s depending on environmental conditions.

\begin{figure*}[t]
	\centering
	\begin{subfigure}{0.45\textwidth}
		\centering
		\includegraphics[width=\linewidth]{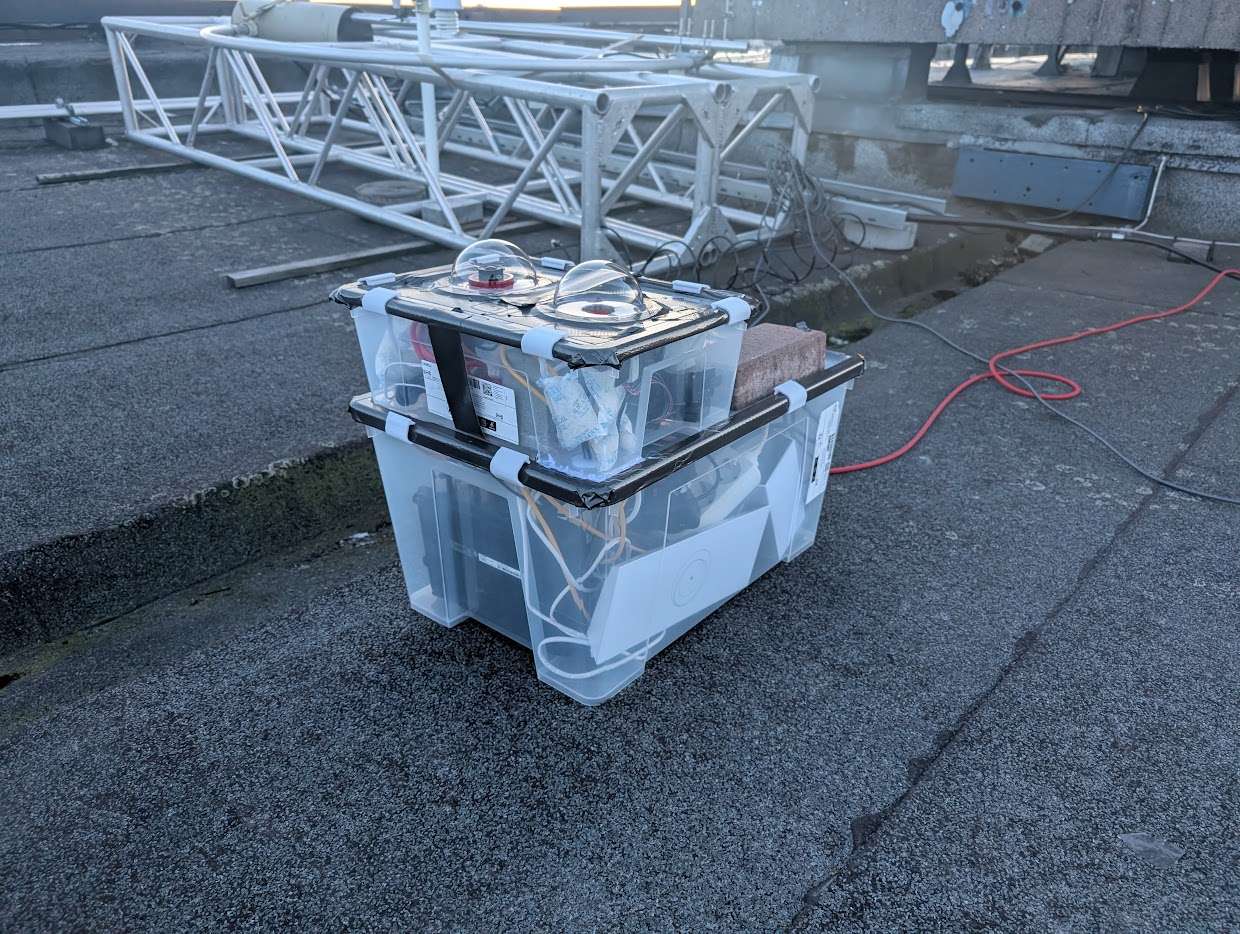}
	\end{subfigure}
	\begin{subfigure}{0.45\textwidth}
		\centering
		\includegraphics[width=\linewidth]{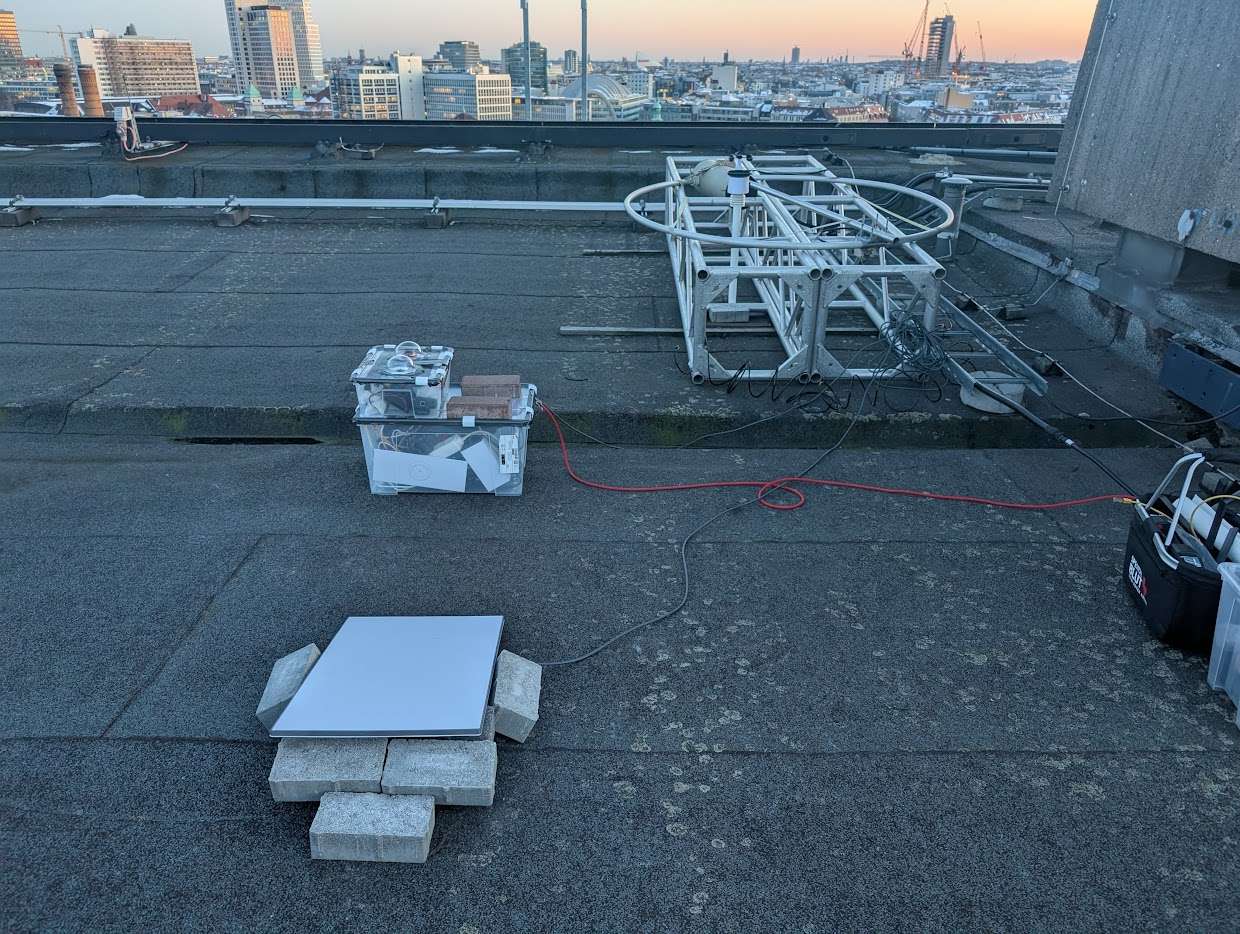}
	\end{subfigure}\\
    \vspace{2mm}
	\begin{subfigure}{0.45\textwidth}
		\centering
		\includegraphics[width=\linewidth]{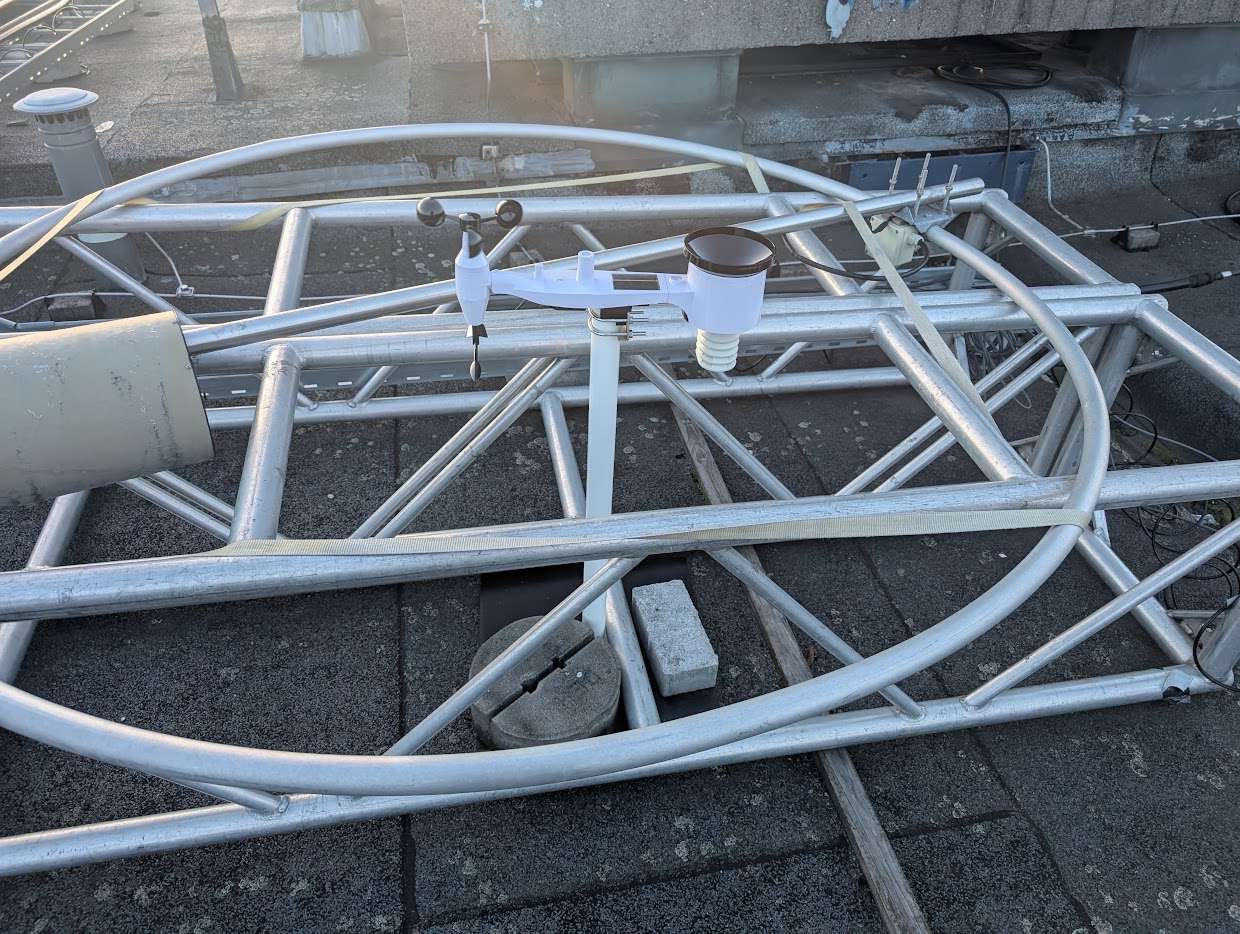}
	\end{subfigure}
	\begin{subfigure}{0.45\textwidth}
		\centering
		\includegraphics[width=\linewidth]{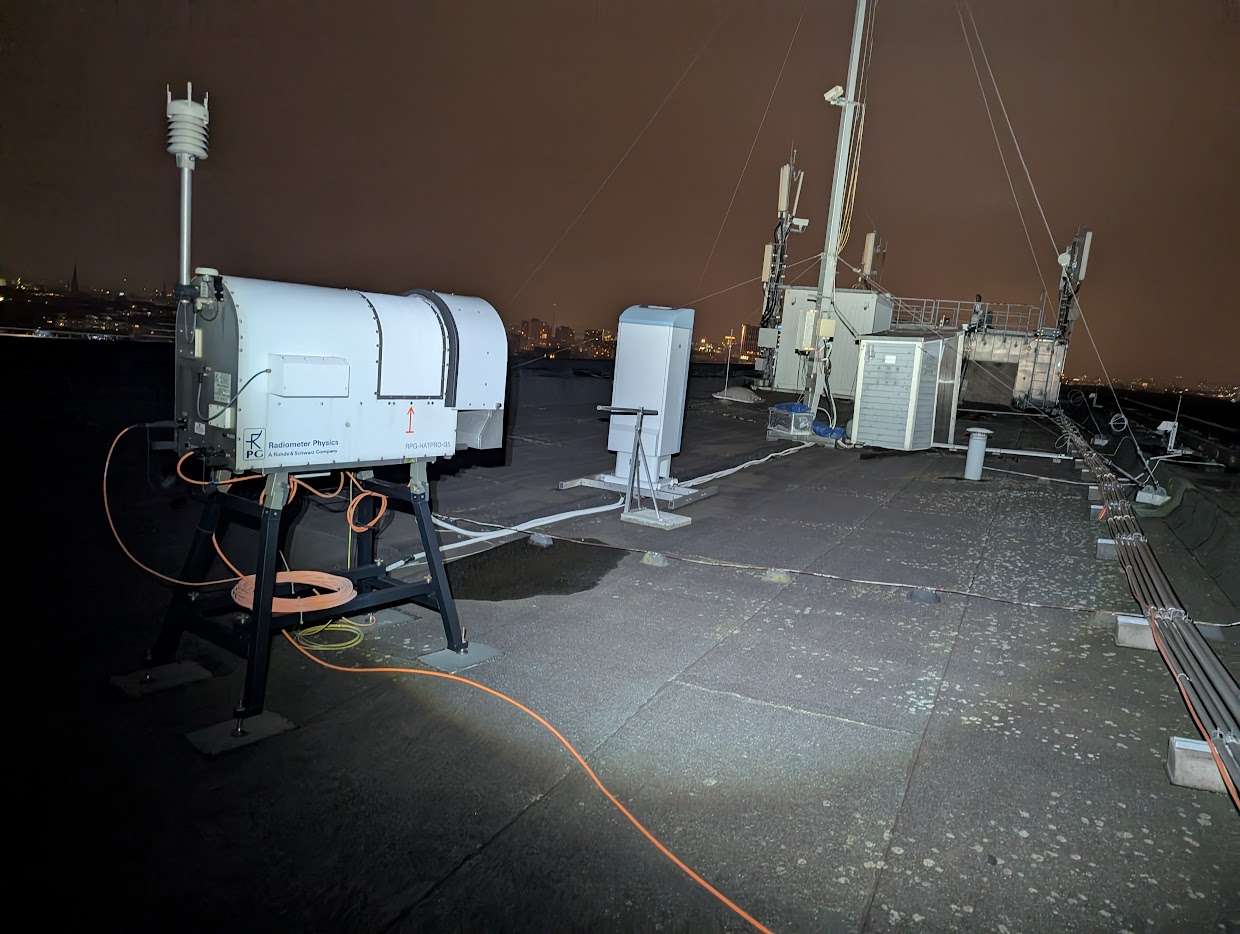}
	\end{subfigure}\vspace{2mm}\\
    \caption{\emph{Upper left:} Close-up photo of stacked transparent plastic boxes containing Starlink measurement equipment. The boxes are placed on a rooftop. 
    \emph{Upper right:} Wide rooftop view showing the Starlink terminal, the stacked box setup, and a nearby weather station. The city skyline is visible in the background. The setup is distributed across the rooftop surface with cables connecting the components.
    \emph{Lower left:} Close-up image of the weather station mounted within a circular metal frame on the rooftop. Various meteorological sensors are visible.
    \emph{Lower right:} Nighttime image of the rooftop showing the microwave radiometer on the left and the ceilometer on the right. The devices are mounted on stands, with cables running along the surface. Background city lights are visible.}
	\label{fig:experiment_setup}
\end{figure*}

In order to ensure reproducibility, we will share our experimental artifacts with the research community together with this paper.

\begin{figure*}[t]%
	\centering
	\begin{subfigure}[b]{0.49\textwidth}
		\includegraphics[width=\textwidth]{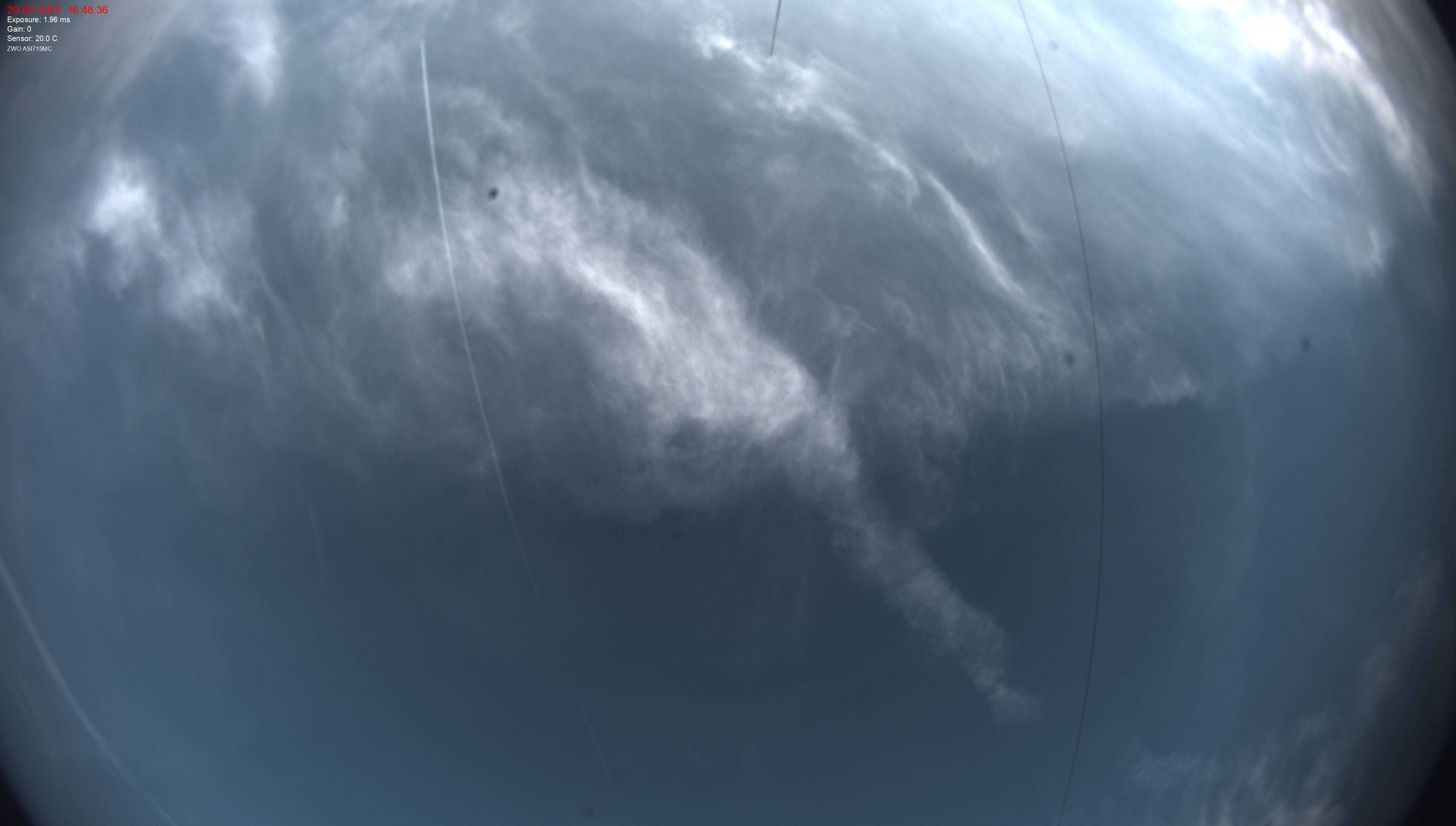}
		\label{fig:day_image}
	\end{subfigure}
        \hfill
	\begin{subfigure}[b]{0.49\textwidth}
		\includegraphics[width=\textwidth]{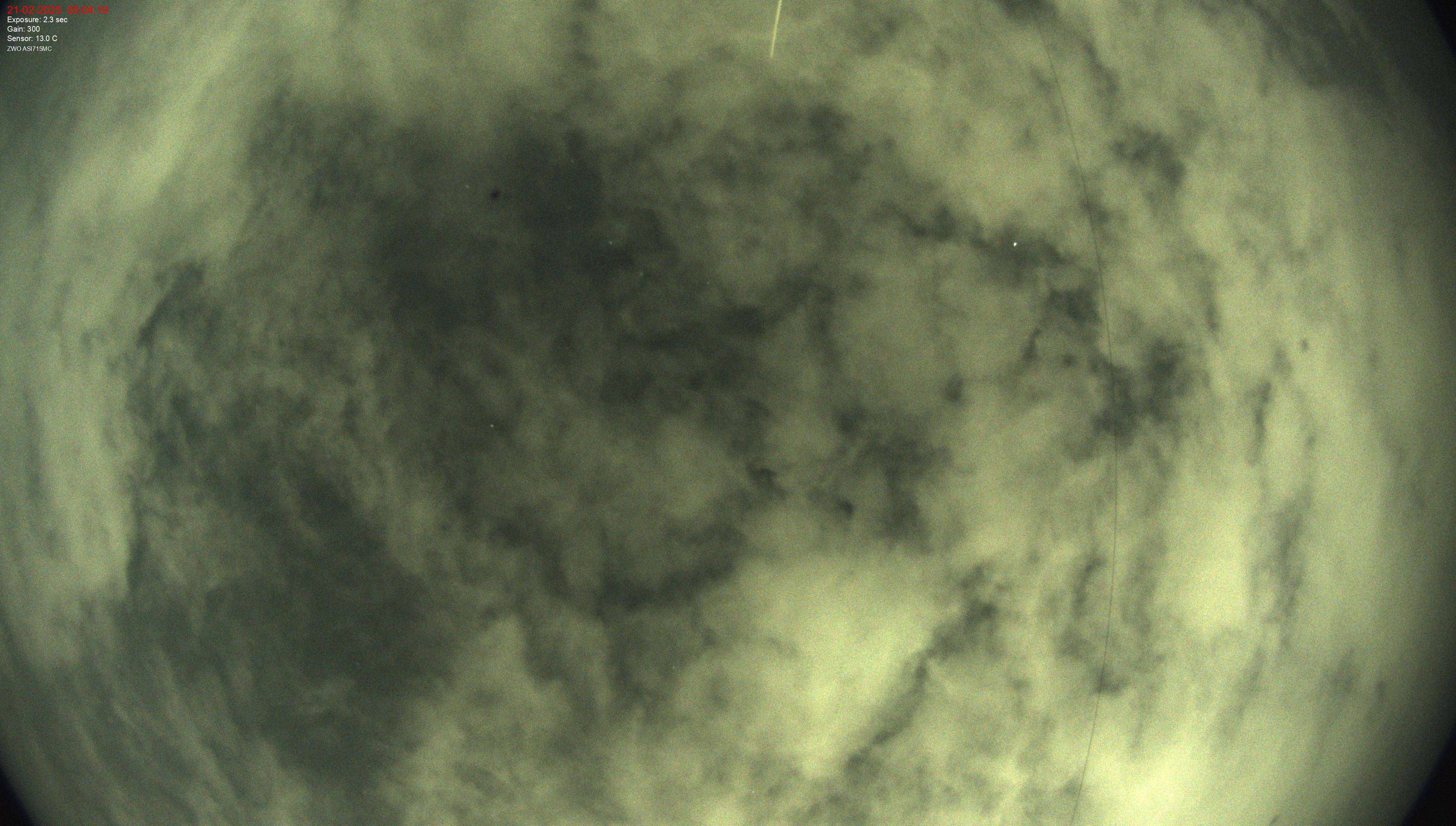}
		\label{fig:night_image}
	\end{subfigure}
        \caption{Two fisheye images of the sky used in cloud classification. The left image shows a daytime sky with sparse clouds, labeled as 3 out of 8 oktas. The right image shows a nighttime sky with dense cloud cover, labeled as 6 out of 8 oktas. These images represent varying lighting and cloud conditions.}
        	\label{fig:day_night_examples}
\end{figure*}

\section{Related Work} \label{sec:related_work}

\subsection*{Performance Measurements of LEO Satellite Networks} \label{sec:related_work_performance_measurements}

Since the start of Starlink’s public service in 2021, numerous studies have characterized LEO network performance. Early work in 2022 analyzed browser-based performance, throughput, and latency \cite{kassemBrowsersideViewStarlink2022, michelFirstLookStarlink2022}, showing clear gains over GEO systems and performance comparable to DSL, with downloads up to 146\,Mbit/s and median RTT around 50\,ms. Subsequent studies examined routing behavior and network characteristics behind the user terminal \cite{maNetworkCharacteristicsLEO2023, panMeasuringLowEarthOrbitSatellite2023}. 

Recent work has expanded measurement methodologies and scale. \cite{izhikevichDemocratizingLEOSatellite2024} introduced the \textit{HitchHiking} inside-out method, enabling global measurements without dedicated dishes and revealing latency anomalies linked to Starlink’s 15-second scheduler and inter-satellite links. Large-scale crowdsourced analyses further highlighted scheduling effects and regional performance differences \cite{mohanMultifacetedLookStarlink2024}. Finally, \cite{panMeasuringSatelliteLinks2024} combined inside-out and outside-in measurements to study one-way delay and RTT between terminals, ground stations, and PoPs.

\subsection*{Impact of Weather Conditions on LEO Satellite Networks} \label{sec:related_work_weather_performance}

Weather-related attenuation has long been studied in satellite communications, mainly for GEO systems, using link-budget models and long-term propagation data \cite{panagopoulosSatelliteCommunicationsKU2004a}. However, these works do not focus on modern large-scale LEO constellations or dense local sensing.

Recent Starlink-focused studies have begun quantifying weather impacts. Prior work detected rain-related throughput degradation \cite{kassemBrowsersideViewStarlink2022, maNetworkCharacteristicsLEO2023}, and \cite{laniewskiWetLinksLargeScaleLongitudinal2024} reported significant download reductions during and around precipitation. A follow-up study used camera-based cloud observations over six months, suggesting throughput reductions above 6\,Mbit/s but noting measurement artifacts and daytime-only coverage \cite{lanferObservingSkiesGroundBased2024}. Additional studies link heavy rain to 2--3\,dB Ku-band attenuation and highlight possible Ka-band cloud effects \cite{vasishtL2D2LowLatency2021}, while early tests indicate temperature sensitivity \cite{maNetworkCharacteristicsLEO2023}.

Our work extends this line of research by combining a broader set of meteorological variables with professional instruments (e.g., ceilometer, microwave radiometer, day-and-night sky camera) and higher-resolution network measurements. This enables a more detailed joint analysis of weather and link dynamics and provides a fine-grained local perspective on LEO link optimization for the latest Starlink generation.

\section{Methodology}\label{sec:methodology}

Throughout the analysis, we report Pearson's correlation coefficient $r$ and associated $p$-values to quantify linear relationships between weather variables and network metrics. Here, $r\in[-1,1]$, where $|r|<0.3$ indicates weak, $0.3\leq|r|<0.5$ moderate, and $|r|\geq0.5$ strong linear correlation.

\subsection{Measurements Overview}\label{sec:measurements-overview}

Deployment began on January 3, 2025, on the university roof, followed by a two-week testing period. The modular, waterproof setup supports power injection and withstands wind speeds up to 160\,km/h and wide temperature ranges.

Basic meteorological variables are recorded by the semi-professional \textit{Ecowitt GW1101:WS69} and professional \textit{Vaisala WXT536} stations, measuring temperature, humidity, solar/UV index, rainfall, and wind. A night-capable sky camera provides image-based cloud estimates.

Advanced instruments from the Department of Climatology include a microwave radiometer (\textit{RPG-HATPRO-G5}), ceilometer (\textit{Lufft CHM15k}), disdrometer (\textit{Thies Clima LNM}), and Eddy-Covariance system, capturing variables such as liquid water path (LWP), water vapor, and cloud properties.

The ceilometer is a lidar-based instrument that detects cloud base height and coverage via backscatter, reporting cloud amount in oktas (0–8). The microwave radiometer retrieves temperature and humidity profiles, integrated water vapor, and LWP, enabling continuous column measurements.

For image-based estimation, an Astro Camera (\textit{ZWO SI715MC Color}) with a 150° lens was installed to match the dish FOV. Controlled by AllSky\foothref{https://github.com/AllskyTeam/allsky}, it captures images every 120\,s using auto-exposure/gain.

On the network side, we measure throughput, RTT, traceroute, Starlink gRPC debug data, obstruction maps, and POP latency (every 10\,ms). Most weather and gRPC data have 60\,s resolution; active network tests and images follow a 2-minute interval.

\Autoref{fig:experiment_setup} shows a photo of our setup.

\subsection{Network Setup}\label{sec:network-setup}

The router runs openWRT with three WAN interfaces:
\begin{enumerate}
	\item \texttt{wanUNI} – University WAN (LAN)
	\item \texttt{wanEduroam} – University WAN (WiFi)
	\item \texttt{wanSAT} – Starlink WAN (LAN)
\end{enumerate}

Priority is given to \texttt{wanUNI} and \texttt{wanEduroam} to isolate benchmark traffic on \texttt{wanSAT}. Tests run directly on the router using \texttt{iperf3}, \texttt{fping}, and \texttt{traceroute}.
%
A static-IP VPS in Central Europe serves as the constant target. UDP is used to avoid TCP congestion effects and measure raw bandwidth. Tests run sequentially every 2 minutes via cron.
Devices are connected through a PoE switch with static IPs. The Starlink router operates as gateway (not in bypass mode) to maintain remote access and enable richer gRPC data and app-based monitoring. Because Starlink uses CG-NAT, a \texttt{Rathole}\foothref{https://github.com/rapiz1/rathole} tunnel provides external access.
To infer connected satellites, we build on \cite{ahangarpourTrajectorybasedServingSatellite2024} and SatInView\foothref{https://github.com/aliahan/SatInView}, extending their scripts to reconstruct suspected satellites and distances for the 56-day period at 1-second granularity.

\subsection{Cloud Cover Classification}\label{sec:cloud_classification}

Cloud cover must be measured with high spatial and temporal alignment to the dish FOV. Using the 0–8 okta scale, we combine ceilometer data, sky images, and microwave radiometer measurements. This provides higher locality than satellite or radar products and remains cost-effective.

\subsubsection{Classification through Vision Language Models}\label{sec:classification-vision-language}

From February 20 to April 10, 2025, 35,325 images were captured. We manually labeled 2,100 images (balanced day/night), with 500 each for validation and testing. Due to artifacts such as lens flare and raindrops, context-aware VLMs were preferred. VLMs enable cloud estimation with the same cadence and FOV as the dish. Prior evaluations rank \texttt{gpt-4o} among the strongest vision models \cite{leeVHELMHolisticEvaluation2024}, and it ranked 13th on the Open VLM Leaderboard (April 2025).

\subsubsection{Input \& Pre-Processing}\label{sec:input-preprocessing}

Images were filtered, resized from 3864×2192 to 512×290 using Lanczos resampling, and \texttt{base64}-encoded for API transmission. 
Example images are shown in \Autoref{fig:day_night_examples}.

\subsubsection{Vision Prompting via OpenAI API}\label{sec:vision-prompting}

The multimodal \code{gpt-4o} model received structured instructions to (1) validate sky images, (2) justify invalidity, (3) estimate cloud cover (0–8), and (4) list artifacts, returning strict JSON. Temperature was set to 0.2 and outputs capped at 500 tokens. Asynchronous processing enabled classification of all images in $\sim$10 hours (ca.~2.5\,s per image).

\subsubsection{Post-Processing \& Result Storage} \label{sec:image_classifier_post_processing}

Responses were parsed into a dataframe and periodically saved to CSV. Of all images, 1,685 (4.8\%) were marked invalid, mainly due to “no sky” (ca.~70\%), water droplets (ca.~15\%), and lens distortion (ca.~10\%).

\subsubsection{Test of Accuracy and Prompt-Engineering} \label{sec:image_classifier_accuracy}

Prompt optimization proceeded over four iterations using 300 training images. The final version achieved an average deviation of 0.61 oktas on the 500-image test set (92.25\% precision). Human annotators differed by 0.38 oktas on average, highlighting inherent subjectivity.

Adding a preliminary validity check reduced error from 0.99 to 0.61 oktas. Invalid images would otherwise yield an average error of 7.43. Remaining errors stem mainly from low light and ambiguous skies. Further fine-tuning with custom GPTs was considered but not pursued due to sufficient zero-shot performance and cost constraints.

\section{Data Preparation}
\subsection{Data Cleaning \& Outage Mitigation}\label{sec:data-cleaning}

During the exploratory phase, a significant and abrupt increase in upload speeds was observed on March 18th, at approximately 22:30. At first, a software update was assumed to be the cause. After the extraction of the software update reboot logs of the dish from the gRPC data, the two closest updates were on March 11th and 20th, and therefore do not stand in temporal connection to the increase.
After rigorously ruling out local setup or hardware modifications as possible causes, Starlink technical support was contacted. The support team confirmed by telephone that an internal infrastructure change occurred at the Point of Presence (POP) during that time.
Consequently, to maintain comparability and data integrity in the subsequent analysis, the mean increase (26.7 MBit/s) in upload speeds observed after this event was subtracted from the affected measurements. The general positive trend of download, upload, and latency was fit in a linear regression (ignoring outliers) and afterwards adjusted to a zero slope and reapplied to the data, effectively removing the general trend during the period for better comparability. The POP Ping latency measurements were also filtered for bandwidth disturbance artifacts and are conducted only for parts of the experiment period.

This shows that Starlink itself is still under heavy development and brings fluctuation in terms of stability and consistency. These adjustments help mitigate the impact of unintentional systematic shifts caused by infrastructure changes outside the study's control. These pre-processing steps ensured that the dataset was both consistent and representative of actual conditions, providing a reliable foundation for analyzing the interactions between Starlink network metrics and weather influences. 

\subsection{Day Drift Correction of Network Variations}\label{sec:day-drift}

During the analysis of the Starlink network throughput data, a systematic diurnal variation of mean throughput was observed, with a pronounced daily drift. Specifically, the mean throughput exhibited a peak during the early morning hours (around 5:00 local time) and a minimum during the evening (around 20:00 local time). This pattern suggests that the network load, and hence available throughput, is subject to significant hourly fluctuations across the day, likely due to aggregated user demand. Comparing the dataset with the same hour only would allow an analysis without the influence of this drift. However, this dramatically decreases the sample size. To enable a meaningful comparison of throughput measurements across different hours and days, it is essential to correct for this diurnal drift. Without such correction, hourly variations in the dataset may confound analyses aimed at isolating other effects or comparing underlying trends. On the other hand, a disadvantage of this is the potential elimination of weather-based diurnal influence. \\

\begin{enumerate}
\item \textbf{Timezone Normalization:} Timestamps were converted from UTC to \texttt{Europe/Berlin} local time.
\item \textbf{Selection of Baseline Days:} Only days with a ceilometer mean cloud cover below 1 (i.e., minimal weather impact) were used to establish a diurnal baseline.
\item \textbf{Hourly Baseline Calculation:} Mean throughput was calculated for each local hour across these low-cloud days.
\item \textbf{Derivation of Correction Offsets:} For each hour, the difference between the global and hourly mean throughput was determined.
\item \textbf{Correction Application:} This offset was added to all measurements, normalizing throughput across all hours.
\end{enumerate}

\begin{figure*}[t]
	\centering
	\begin{subfigure}{0.49\textwidth}
		\centering
		\includegraphics[width=\linewidth]{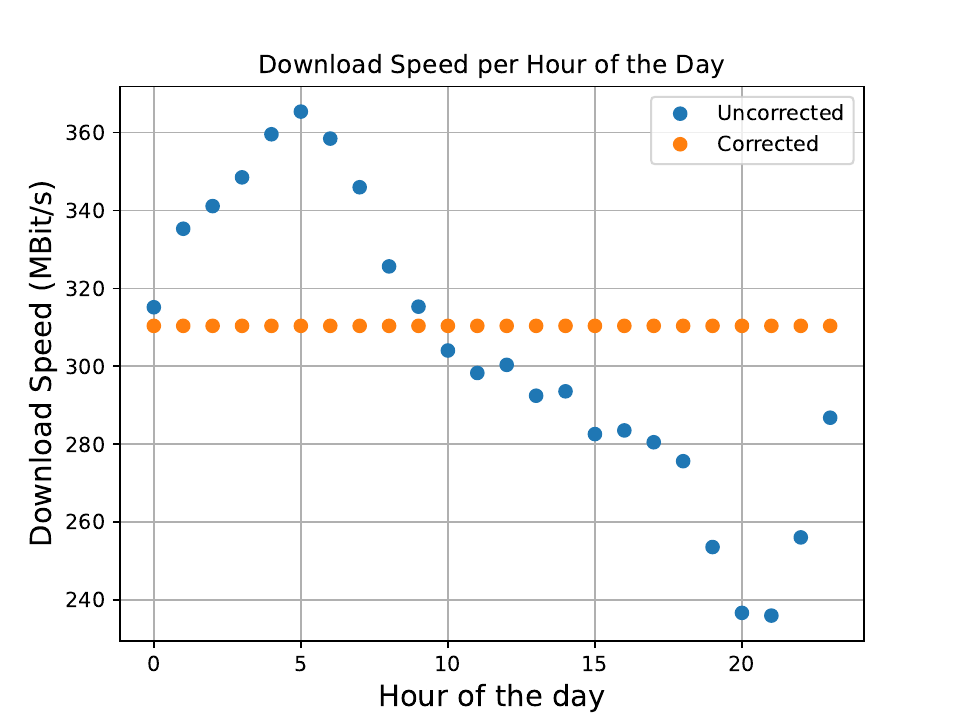}
	\end{subfigure}%
	\hfill
	\begin{subfigure}{0.49\textwidth}
		\centering
		\includegraphics[width=\linewidth]{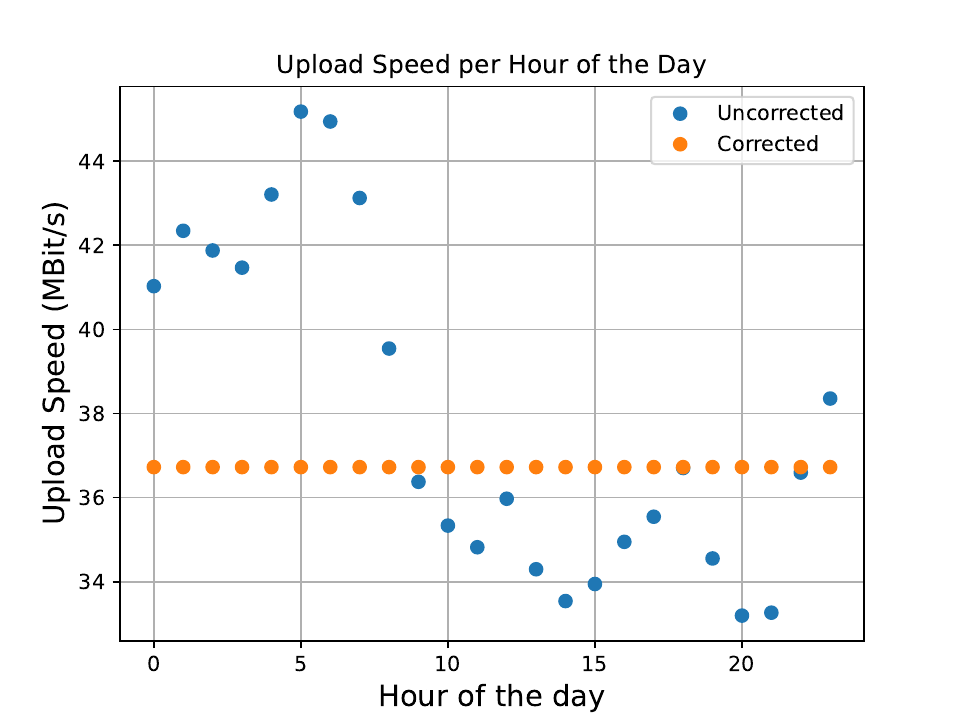}
	\end{subfigure}
	\label{fig:day_drift_correction}
        \caption{Day drift correction. Two scatter plots comparing uncorrected and corrected throughput values by hour of the day. The left panel shows download speed in MBit/s, where uncorrected values fluctuate by hour, while corrected values form a flat horizontal line. The right panel shows upload speed with a similar pattern: scattered uncorrected values and a constant corrected baseline.}
\end{figure*}

Days with the least weather impact were used to reflect maximum possible values. Correction was applied to the \texttt{Download}, \texttt{Upload}, and \texttt{Ping} datasets. The specific choice of target mean is not critical, provided all values are adjusted consistently.

\subsection{Merging \& Resampling}\label{sec:merging-resampling}
To facilitate integrated analysis, the datasets were merged primarily based on their timestamp indices using a nearest-neighbor approach with a tolerance of five minutes. This ensured that measurements from different systems were temporally aligned as closely as possible, preserving data consistency for subsequent analysis. The five-minute tolerance was chosen to prevent merging measurements with higher temporal differences due to data measurement gaps.

Given that most measurements were originally sampled at high frequency, typically every 1-2 minutes, the resulting merged dataset contained a large number of closely spaced observations. For the purposes of exploratory analysis, plotting, and correlation studies, the data was resampled to a coarser, one-hour interval using appropriate aggregations (such as mean values).

This resampling process is justified since weather conditions and their impact on network performance tend to vary over longer timescales compared to the rapid two-minute sampling interval. Additionally, measurements taken by the ceilometer and microwave radiometer only capture conditions in a straight line above the instrument due to their laser functionality. Valid measurements over a larger area of the sky can therefore only be achieved through temporal integration. Aggregating the data into hourly windows overcomes these limitations while also reducing short-term noise and variability, allowing for clearer identification of broader trends and dependencies. This approach enhances interpretability without sacrificing essential information relevant to the research questions. Therefore, all the following data and figures are resampled to a 1-hour window [1h RS] and averaged using the arithmetic mean.

\section{Results} \label{sec:results}
\subsection{Overall Network Performance} \label{sec:network_performance}
Visual inspection of the main network metrics is the first step of data preparation.

We measured the average download speed per day throughout the whole experiment period. While the throughput is noisy, a global upward trend can also be seen. An outlier on the 30th of March can be explained by a very rainy day, which heavily reduced the download speed. 

\begin{figure*}[t]
    \centering
    \includegraphics[width=.9\linewidth]{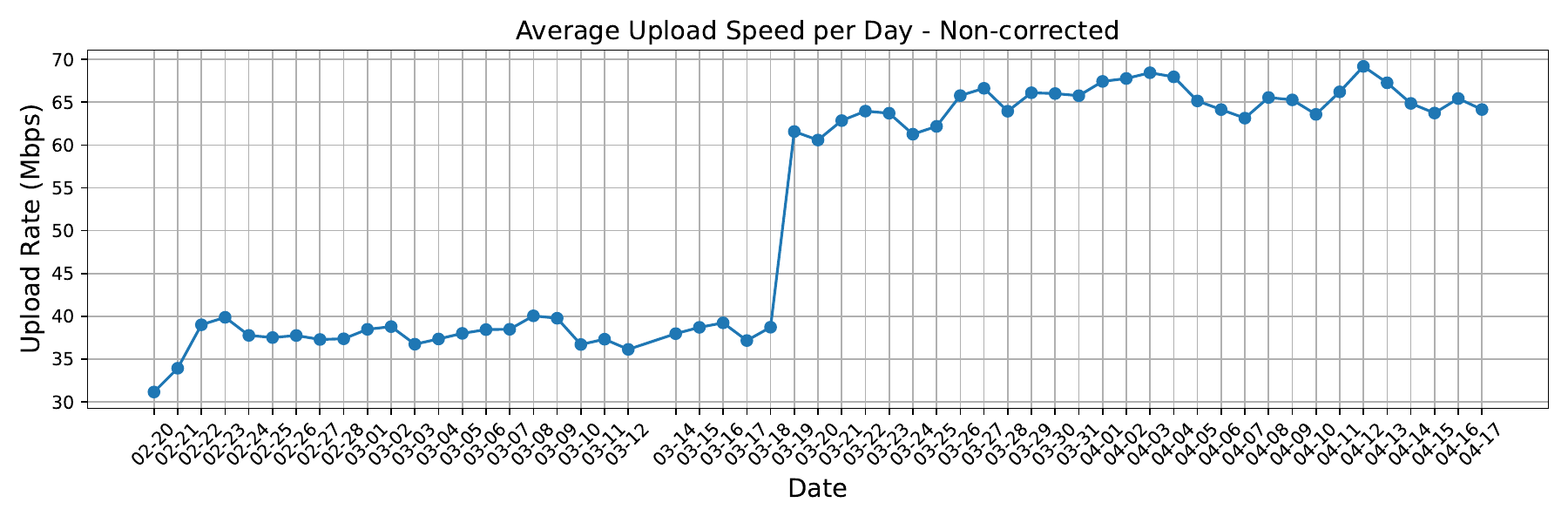}
    \caption{Raw upload throughput per day. Line plot showing the average upload speed per day over a period from early May to mid-June. The y-axis represents upload rate in MBit/s, and the x-axis shows dates. The upload speed is relatively flat around 35–40 MBit/s until mid-May, then sharply increases to around 60–70 MBit/s and remains stable.}
    \label{fig:raw-upload}
\end{figure*}

The upload throughput (\Autoref{fig:raw-upload}) remains more stable when observed in small windows and therefore undergoes less fluctuation. However, a large increase in average bandwidth is seen beginning on the 19th of March. Besides this rapid increase, the graph additionally indicates an overall positive slope. 
The ping latency also undergoes variations of up to 7ms on a day average. In particular, three days at the end of March show a significant increase of 6ms with a following sudden drop to the original value afterwards. All network measurements have a gap on the 13th of March, which is caused by an electricity cut due to maintenance operations on the main roof done by the administration. For a thorough analysis of these metrics, a correction for general drifts, outliers, and trends is crucial, which is explained in the following sections.

The low-weather-impacted days were chosen to adjust the values to the maximum value possible when there is no weather impact. Another approach would have been to use the overall mean of every day as a target value to get a more realistic correction. We argue that the target value is not crucial as long as all values are adjusted to the same mean.
The calculated correction is applied to the \texttt{Download, Upload, Ping} datasets.

\subsection{Starlink Reconfiguration Schedule} \label{sec:starlink_routing_schedule}
\begin{figure*}[t]%
	\centering%
	\includegraphics[width=0.99\textwidth]{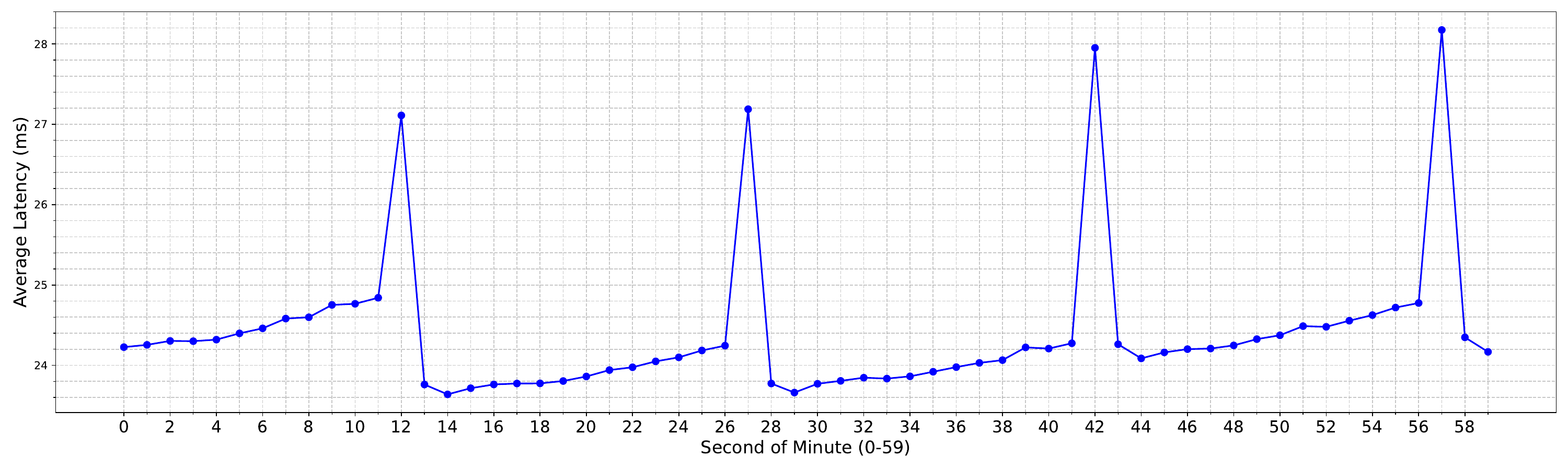}%
        \caption{Average latency by second of minute. Line plot showing average latency in milliseconds by second of the minute (0–59). Latency is relatively low and stable but shows sharp spikes at seconds 12, 27, 42, and 57. This indicates periodic latency increases at 15-second intervals.}
	\label{fig:latency_interval}%
\end{figure*}

Similar to other papers \cite{tanveerMakingSenseConstellations2023,ahangarpourTrajectorybasedServingSatellite2024,izhikevichDemocratizingLEOSatellite2024, panMeasuringSatelliteLinks2024}, our experiment shows the same spikes in latency for the global schedule interval of Starlink, which occurs exactly every 15 seconds at seconds 12, 27, 42, 57 of every minute. At these timestamps, the Starlink network has a globally synchronized reconfiguration period, which leads to a change of the currently connected satellite in most cases to ensure a well-performing link. Our measurements include ping tests every 10ms to the Starlink default gateway \texttt{100.64.0.1}, which is always the first network hop after leaving our local network according to the traceroute logs. These measurements for every second result in clearly visible intervals as shown in \Autoref{fig:latency_interval}.

\subsection{Investigation of Serving Satellite Data} \label{sec:investigation_serving_satellite_data}
The 60-second interval can be split into four sub-intervals
\[
T_1 = [12, 27) \quad
T_2 = [27, 42) \quad
T_3 = [42, 57) \quad
T_4 = [57, 12)
\]
\\
At the start of each interval, the latency increases by 3.1 milliseconds (13\%) on average. However, we observe slight differences in the interval latencies. While the pre-reconfiguration latencies of $T_2$ and $T_3$ are at 24.2ms, those of $T_1$ and $T_4$ are at 23.8.

After extracting the satellite data as described in the previous section, the data were subsequently grouped by the connected satellite. This allows us to investigate the influence of different parameters, such as the connected satellite version and the exact distance to the user terminal. Furthermore, the data was grouped by the count of consecutive connections to the same satellite. Usually, a new satellite is selected for the active connection at every reconfiguration (4 times per minute), a so-called satellite handover. However, occasionally the dish may connect to the same satellite again, e.g., in cases where there is no better satellite in reach.

Specifically, the satellite is changed every 1.8 intervals, or 2.2 times per minute on average. When calculating the mean distance per consecutive satellite, we see, for logical reasons, also an increasing average distance to the satellite, as shown in \Autoref{tab:consecutive-latency} (count of 1 means the satellite is only connected for 1 interval). The maximum of consecutive intervals connected to the same satellite was observed once, with 10 intervals. This means the dish had an established link to this satellite for 2.5 minutes. The latency spikes at reconfiguration seconds occur even if the same satellite is chosen again as the active link. A positive correlation between Consecutive Count and Mean Latency becomes clear, resulting in higher latencies with rising consecutive counts. Assertions about throughput cannot be made because these tests are conducted only every 2 minutes and therefore do not fulfill the detailed requirements in high-resolution, as opposed to the POP ping tests, which are executed every 10ms.


\begin{table}[t]
    \centering
    \caption{Average Distance and Ping Latency grouped by Satellite Consecutive Count}
    \begin{tabular}{c c c}
        \toprule
        \textbf{Count} & \textbf{Distance (km)} & \textbf{Latency (ms)} \\
        \midrule
        \textbf{1} & 577.27 & 27.63 \\
        \textbf{2} & 559.49 & 27.50 \\
        \textbf{3} & 560.56 & 27.51 \\
        \textbf{4} & 572.35 & 27.69 \\
        \textbf{5} & 588.45 & 28.06 \\
        \textbf{6} & 607.92 & 28.51 \\
        \textbf{7} & 628.68 & 29.40 \\
        \textbf{8} & 651.62 & 29.95 \\
        \textbf{9} & 691.06 & 32.12 \\
        \bottomrule
    \end{tabular}
    \label{tab:consecutive-latency}
\end{table}

When grouping by Starlink Satellite Version, a significant increase in average Download and Upload speed is observed for the latest version \textit{v2-mini}. The new version brings an improvement of approximately 34 MBit/s in download, 2 MBit/s in upload, and 1.4ms in latency, while also being in a lower Earth orbit (cf. \autoref{tab:version-performance}).

\begin{table}[t]
    \centering
    \caption{Network Performance by Starlink Satellite Version}
    \begin{tabular}{l r r r}
        \toprule
        \textbf{Metric} & \textbf{v1.0} & \textbf{v1.5} & \textbf{v2-Mini} \\
        \midrule
        Latency (ms)        & 25.26 & 25.29 & 23.82 \\
        Download (MBit/s)   & 296.38 & 294.36 & 326.82 \\
        Upload (MBit/s)     & 35.51 & 35.49 & 38.07 \\
        Distance (km)       & 610.99 & 627.79 & 533.27 \\
        \bottomrule
    \end{tabular}
    \label{tab:version-performance}
\end{table}

\subsection{Precipitation}\label{sec:precipitation}\label{sec:weather_impact}

\Autoref{tab:distrometer-precip-corr} shows a weak to moderate negative correlation between download speed and precipitation intensity ($r = -0.25,$ $ p < 0.001$). While the effect size is modest, the low p-value indicates that the relationship is statistically robust. This suggests that increased precipitation is typically associated with reduced download speeds, while upload speeds and latency remain largely unaffected.

\begin{figure}[t]
	\centering
	\includegraphics[width=0.95\linewidth]{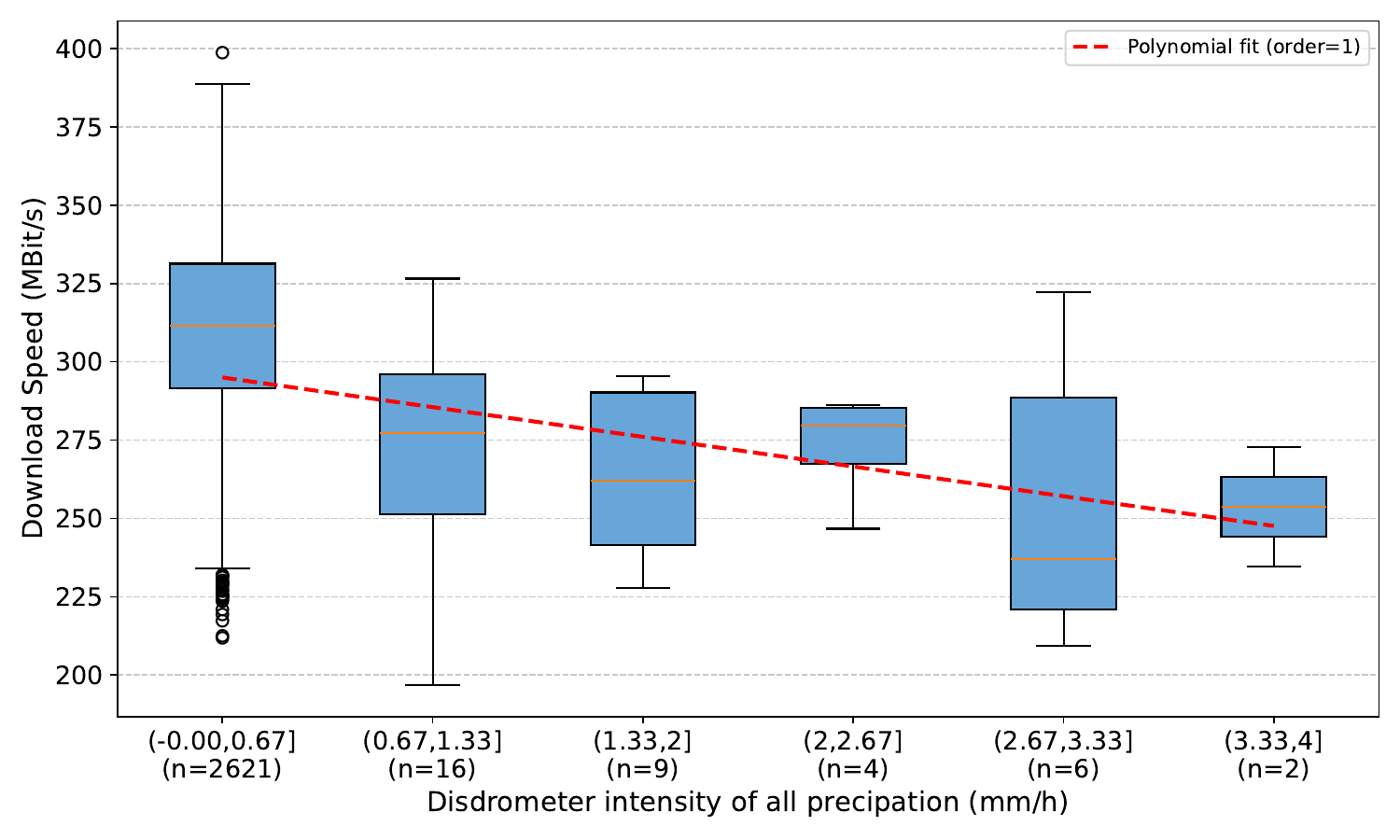}\\
        \caption{
        Boxplot showing download speed in MBit/s versus precipitation intensity (in mm/h) based on 30-minute RS intervals. The x-axis is divided into six precipitation bins ranging from 0 to 4 mm/h. The median download speed decreases as precipitation intensity increases. A red dashed line represents a polynomial fit, showing a downward trend.}	\label{fig:download_rain}
\end{figure}

\Autoref{fig:download_rain} shows the download throughput versus intensity of precipitation resampled to 30min, being a suitable interval for rain events. The drop in download speed is visible, although the sample size for the rain bins is not very high due to relatively rare rain events during the experiment period. While the median download throughput during non-precipitation periods is around 310 MBit/s, it drops to around 250 MBit/s during heavy rain. During single heavy rain events, the download bandwidth dropped to around 108 MBit/s.

The observed decrease in throughput during precipitation can be explained by the effect of rain on the propagation of satellite signals. Rain causes attenuation and scattering of radio waves, especially at the frequencies used by Starlink, which leads to a reduction in signal strength as it passes through areas of heavy precipitation. This phenomenon can result in lower data rates and increased error rates, thus explaining the reduction in download speeds seen during rain events.

\begin{table}[t]
	\centering
        \caption{Correlation of Network Metrics vs. Disdrometer 1M Intensity of All Precipitation (Rain Only)}
	\begin{tabular}{l|c|c|c}
		\toprule
		\textbf{Metric} & \textbf{Sample Size} & \textbf{r Value} & \textbf{p Value}       \\
		\midrule
		Download Speed          & 1330                 & -0.25            & $1.39 \times 10^{-20}$ \\
		Upload Speed            & 1330                 & -0.05            & 0.070                  \\
		Average Ping            & 1330                 & 0.02             & 0.452                  \\
		\bottomrule
	\end{tabular}
	\label{tab:distrometer-precip-corr}
\end{table}

\subsection{Liquid Water Path (LWP) \& Integrated Water Vapor (IWV)}\label{sec:lwp}

Liquid Water Path (LWP) quantifies the total amount of liquid water in a vertical column of the atmosphere, measured in grams per square meter ($g/m^2$). Higher LWP values are indicative of thicker clouds or increased atmospheric moisture.

\begin{figure}[t]
	\centering
	\includegraphics[width=0.95\linewidth]{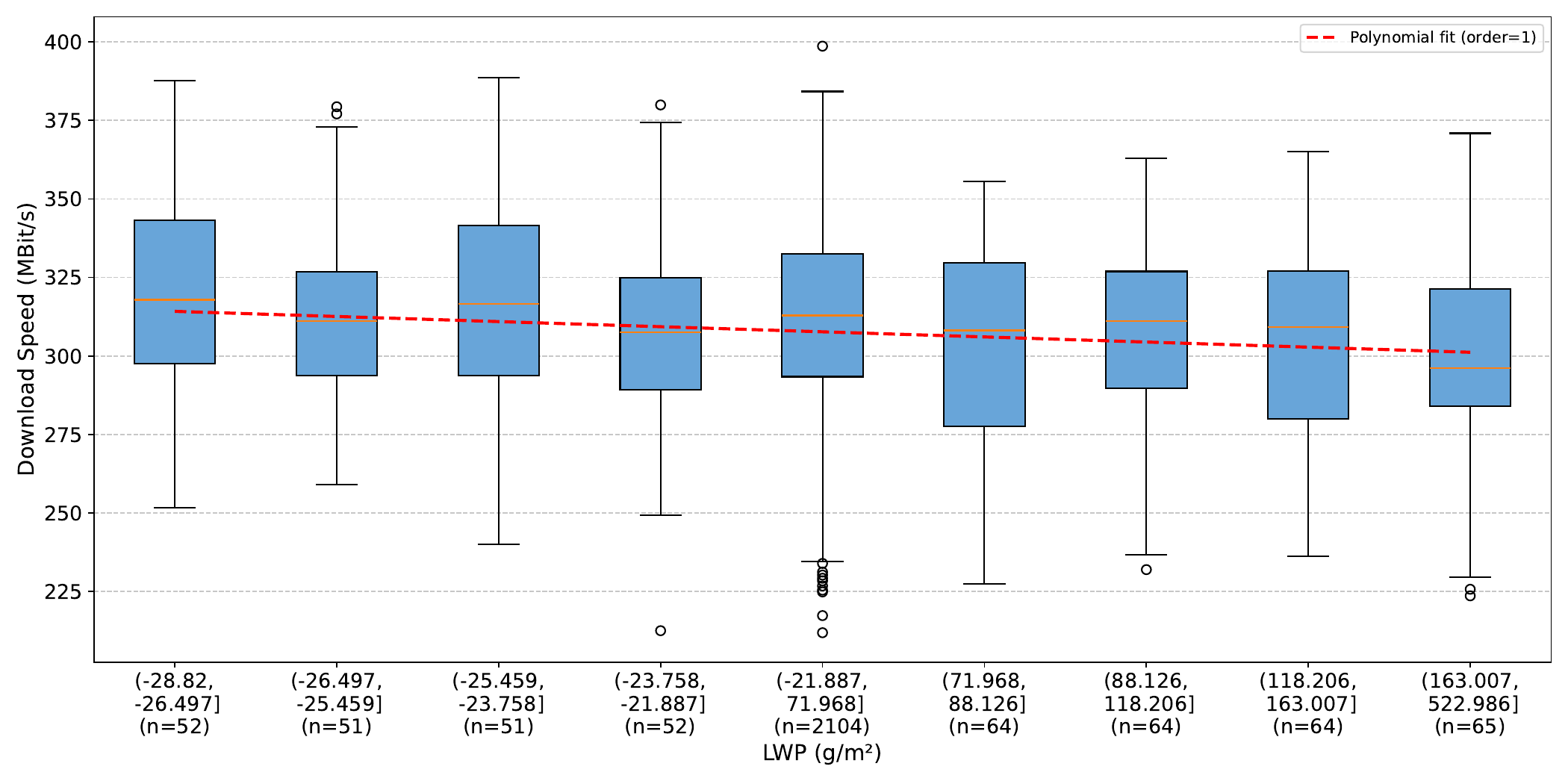}
        \caption{Boxplot of download speed in Mbps versus liquid water path (LWP in g/m²), filtered to exclude rain, using 1-hour RS intervals. The x-axis contains 10 bins of increasing LWP values. A slight downward trend is visible across bins. A red dashed line indicates a polynomial fit, which also shows the downward trend.}
	\label{fig:download_lwp}
\end{figure}

\Autoref{fig:download_lwp} shows the relationship between download speed and LWP with rain events filtered out, revealing a downward trend with increasing LWP. \Autoref{tab:lwp-corr-rain} and \Autoref{tab:lwp-corr-norain} provide the corresponding correlation coefficients: during rain, there is a negative correlation between download speed and LWP ($r=-0.28, p<0.001$), and without rain, the correlation is weaker but still significant ($r=-0.15, p<0.001$). Upload speed and average ping do not show significant correlations with LWP, as reflected by r values near zero and high p values. The median download throughput falls from 320 MBit/s with very few water droplets to 290 MBit/s.

The negative correlation suggests that higher concentrations of atmospheric liquid water are associated with reduced download speeds. This can be explained by additional signal attenuation and scattering as radio waves pass through denser cloud layers, especially under rainy conditions. This shows that the impact of clouds measured by their amount of water droplets has a significant effect on the network performance, even when there are no rain events.

\begin{table}[t]
    \centering
    \caption{Correlation of Network Metrics vs.\ LWP (With Rain)}
    \resizebox{\columnwidth}{!}{
    \begin{tabular}{l c c c}
        \toprule
        \textbf{Metric} & \textbf{Sample Size} & \textbf{r Value} & \textbf{p Value} \\
        \midrule
        Download Speed & 1327 & -0.28 & $4.80\times10^{-25}$ \\
        Upload Speed   & 1327 & -0.06 & 0.0235               \\
        Average Ping   & 1327 & 0.07  & 0.00710              \\
        \bottomrule
    \end{tabular}
    }
    \label{tab:lwp-corr-rain}
\end{table}

\begin{table}[t]
    \centering
    \caption{Correlation of Network Metrics vs.\ LWP (Without Rain)}
    \resizebox{\columnwidth}{!}{
    \begin{tabular}{l c c c}
        \toprule
        \textbf{Metric} & \textbf{Sample Size} & \textbf{r Value} & \textbf{p Value} \\
        \midrule
        Download Speed & 1313 & -0.15 & $5.46\times10^{-8}$ \\
        Upload Speed   & 1313 & -0.02 & 0.372               \\
        Average Ping   & 1313 & 0.08  & 0.00343             \\
        \bottomrule
    \end{tabular}
    }
    \label{tab:lwp-corr-norain}
\end{table}

Integrated Water Vapor (IWV) measures the total amount of water vapor in a vertical column of the atmosphere, expressed in kilograms per square meter ($kg/m^2$). It represents the amount of water vapor, which means it quantifies the gaseous phase of water, rather than condensed liquid water as in Liquid Water Path (LWP). Therefore, IWV provides information about the total atmospheric moisture content.

Analysis of the relationship between download speed and IWV reveals a weak negative correlation both with and without rain. The download speed correlates negatively with IWV ($r=-0.13$ with rain and $r=-0.11$ without rain).

Upload speed does not exhibit significant correlations with IWV, with r values close to zero and high p values. Interestingly, the average ping seems to slightly go down with increased IWV. This effect is not very strong with weak r-values and is also challenging to explain. These results indicate that water in its vaporized form has little impact on signal attenuation as opposed to water droplets in the fluid phase.

\subsection{Cloud Cover Detection}
\subsubsection{Ceilometer Cloud Cover Detection}
\Autoref{tab:ceilometer-corr-rain} and \Autoref{tab:ceilometer-corr-norain} show a weak negative correlation between download speed and cloud amount, with $r=-0.14$ with rain and $r=-0.11$ without rain events. In contrast, upload speed and average ping show negligible or no correlation to cloud amount, as reflected by $r$ values around zero and high $p$ values.

These results indicate that increased cloud amount, measured by the ceilometer, may be weakly associated with reduced download throughput. This can be explained by additional attenuation or scattering of the signal caused by dense cloud cover, although the overall impact appears limited and the effect on the median download throughput is only around 10 MBit/s in difference when comparing cloud amount 0 with 315 MBit/s to cloud amount 8 with 305 MBit/s on average.

The ceilometer measurements occasionally also include a vertical visibility value, which estimates the visible distance. With increasing vertical visibility, the ping latency decreases ($r=-0.24, p < 0.0001$).

\begin{table}[t]
    \centering
    \caption{Correlation of Network Metrics vs.\ Cloud Amount (With Rain)}
    \resizebox{\columnwidth}{!}{
    \begin{tabular}{l c c c}
        \toprule
        \textbf{Metric} & \textbf{Sample Size} & \textbf{r Value} & \textbf{p Value} \\
        \midrule
        Download Speed & 1302 & -0.14 & $2.62\times10^{-7}$ \\
        Upload Speed   & 1302 & 0.01  & 0.743               \\
        Average Ping   & 1302 & -0.00 & 0.906               \\
        \bottomrule
    \end{tabular}
    }
    \label{tab:ceilometer-corr-rain}
\end{table}

\begin{table}[t]
    \centering
    \caption{Correlation of Network Metrics vs.\ Cloud Amount (Without Rain)}
    \resizebox{\columnwidth}{!}{
    \begin{tabular}{l c c c}
        \toprule
        \textbf{Metric} & \textbf{Sample Size} & \textbf{r Value} & \textbf{p Value} \\
        \midrule
        Download Speed & 1288 & -0.11 & $8.34\times10^{-5}$ \\
        Upload Speed   & 1288 & 0.03  & 0.344               \\
        Average Ping   & 1288 & -0.01 & 0.617               \\
        \bottomrule
    \end{tabular}
    }
    \label{tab:ceilometer-corr-norain}
\end{table}

\subsubsection{Image-based Cloud Cover Detection}
\begin{figure}[t]%
	\centering%
	%
	\includegraphics[width=0.5\textwidth]{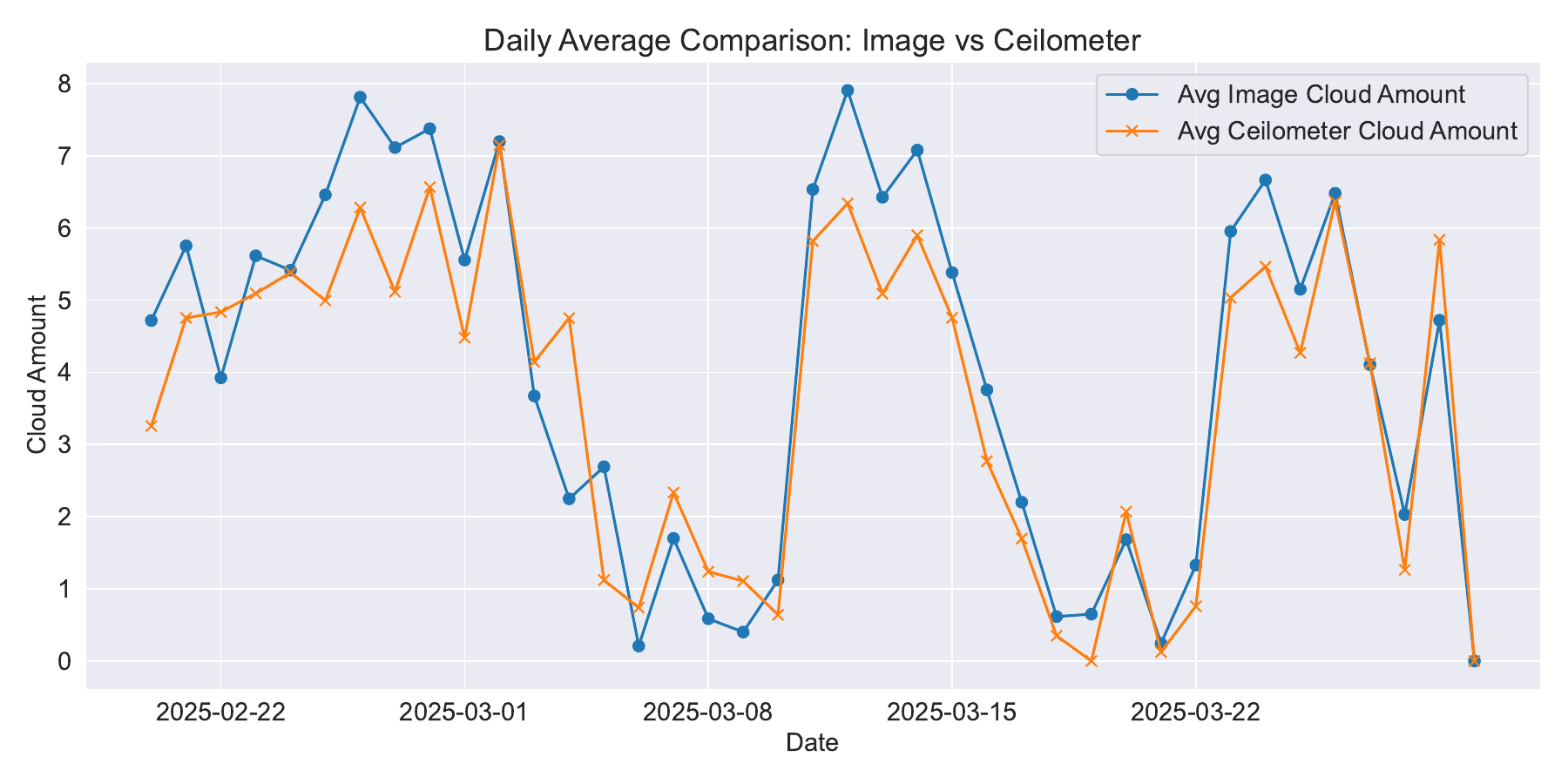}%
        \caption{Line graph comparing daily average cloud amount between an image classifier and a ceilometer from February 22 to March 24, 2025. The X-axis shows the date. The Y-axis shows cloud amount values from 0 to 8. Two lines are plotted: one for average image cloud amount and one for average ceilometer cloud amount. Both lines show similar trends with occasional divergence, particularly during early March.}\label{fig:image_classifier_vs_ceilometer_daily_avg}%
\end{figure}

As mentioned in \Autoref{sec:methodology}, using the test set an average difference between the manual labeling and the classifier of 0.61 [relative: 7.63\%] is achieved. To evaluate the overall performance of the image classifier, it is necessary to consider not only its performance compared to the ground truth, i.e., the manual labeling, but also its significance for cloud measurement. The average deviation in cloud amount between the image classifier and the ceilometer is 1.63 [relative: 20.38\%]. Since the ceilometer can only measure a small area of the sky and only outputs values every 5 minutes, resampling of 1 hour is performed during data evaluation, resulting in an average difference of 1.40 [relative: 17.5\%]. This is further illustrated in \Autoref{fig:image_classifier_vs_ceilometer_daily_avg}, which shows the daily difference between the ceilometer and image classifier. 
Additional experiments show that the ceilometer detects more subtle changes in the clouds than the image classifier, and that the image classifier can primarily be regarded as the upper boundary.

Utilizing Pearson's correlation coefficient, a statistically significant correlation was found between the predicted and reference values, with an $r$-value of 0.78 and a corresponding $p$-value of $3.45 \times 10^{-233}$. In our interpretation, this falls into the range of a strong linear association and confirms the validity of the outputs produced by the image classifier. The Sky Camera used with the Image Classifier provided therefore offers a sufficient approximation to the ceilometer and can thus represent a cost-effective alternative for further investigations. 
\\

\section{Conclusion}\label{conclusion}

This paper analyzed dynamic factors affecting LEO satellite link quality using Starlink as a case study, combining high-resolution network measurements with dense, co-located meteorological sensing. We find that atmospheric liquid water is a key driver of performance variation: rain and elevated liquid water path (LWP) are associated with noticeable reductions in download throughput, while upload and latency are less affected. In contrast, simple cloud presence and integrated water vapor show little explanatory power. We also observe that link quality is shaped by system dynamics such as satellite handovers and infrastructure upgrades, which contribute substantial variability but also measurable performance gains over time.

Future work should broaden the spatial and temporal scope of measurements to other climates, seasons, and regions, and examine additional precipitation types such as snow or hail. Our results further suggest opportunities for weather-aware network management and applications that anticipate moderate, liquid-water-related throughput degradation.

\section*{Acknowledgments}

Many people contributed to this project, and we would like to express our sincere gratitude for their support along the way.
We would especially like to thank the AFuTUB – Amateurfunkgruppe der Technischen Universität Berlin for generously providing access to the experimental site on the TU Berlin main building roof. Their support in planning and deploying the setup, as well as assistance with their local network infrastructure and troubleshooting, was invaluable.
We are grateful to Fred Meier from the Chair of Climatology at the Institute of Ecology (TU Berlin) for providing weather instrument data and for his guidance in interpreting weather-related information relevant to the study.
Our thanks also go to Sebastian Lange from the Institute of Aeronautics and Astronautics of TU Berlin, who made an alternative test environment available and facilitated initial experiments on the roof of the Institute of Aeronautics and Astronautics at TU Berlin.
Lastly, we would like to thank Dominic Laniewski and Eric Lanfer from the University of Osnabrück for sharing their experiences and insights gained from their own Starlink experiment, which served as a valuable starting point for this work.

{

}

\end{document}